\documentclass[aps,prl,twocolumn,superscriptaddress,showpacs]{revtex4}
\usepackage{amsmath,amssymb}
\usepackage[dvips]{graphicx}

\usepackage{dcolumn}
\usepackage{bm}
\usepackage[dvips]{color}

\newcommand{\PR}[4]{Phys. Rev. #1 {\bf #2}, #3 (#4)}
\newcommand{\PRRC}[4]{Phys. Rev. #1 {\bf #2}, #3(R) (#4)}
\newcommand{\PRL}[3]{Phys. Rev. Lett. {\bf #1}, #2 (#3)}
\newcommand{\RMP}[3]{Rev. Mod. Phys. {\bf #1}, #2 (#3)}

\newcommand{\Nature}[3]{Nature {\bf #1}, #2 (#3)}

\def\rnum#1{\expandafter{%
\romannumeral #1}}
\def\Rnum#1{\uppercase\expandafter{%
\romannumeral #1}}

\newcommand{\bol}[1]{\boldsymbol #1}

\begin{document}
\title{Floquet Majorana Edge Mode and Non-Abelian Anyons in a Driven Kitaev Model
}
\author{Masahiro Sato}
\affiliation{Department of Physics and Mathematics, Aoyama-Gakuin University, 
Sagamihara, Kanagawa 229-8558, Japan}
\author{Yuki Sasaki}
\affiliation{Department of Physics and Mathematics, Aoyama-Gakuin University, 
Sagamihara, Kanagawa 229-8558, Japan}
\author{Takashi Oka}
\affiliation{Department of Applied Physics, University of Tokyo, 
Hongo 7-3-1, Bunkyo, Tokyo 113-8656, Japan}

\date{\today}

\begin{abstract}
We theoretically study laser driven nonequilibrium states 
in the Kitaev honeycomb model with a magnetoelectric cross coupling. 
We show that a topological spin liquid with a gapless chiral edge mode emerges 
when we apply an elliptically or circularly polarized laser. 
This is a strongly correlated quantum spin version of the Floquet 
topological insulator. In the topological phase, the edge mode is made from 
Majorana fermions and the bulk has gapped non-Abelian anyon excitations.
%exhcange striction type
\end{abstract}

%75.10.Jm : Quantized spin models, including quantum spin frustration 
%75.40.Gb : Dynamic properties 
%(dynamic susceptibility, spin waves, spin diffusion, dynamic scaling, etc.)
%42.50.Dv : Quantum state engineering and measurements
%75.10.Kt : Quantum spin liquids, valence bond phases and related phenomena
%75.85.+t : Magnetoelectric effects, multiferroics 
%85.60.-q : Optoelectronic devices

\maketitle

%%%%%%%%%%%%%%%%%%%%%%%%%%%%%%%%%%%%%%%%%%%%%%%%%%%%%%%%%%%%
%%%%%%%%%%%%%%%%%%%%%%%%%%%%%%%%%%%%%%%%%%%%%%%%%%%%%%%%%%%%
%%%%%%%%%%%%%%%%%%%%%%%%%%%%%%%%%%%%%%%%%%%%%%%%%%%%%%%%%%%%
%\section{Introduction}
{\it Introduction $-$}
Ultrafast manipulation of quantum systems by laser 
is becoming a hot topic 
in condensed matter~\cite{Kimer05,Kirilyuk10,Karch10,Karch11,Gedik13,Gedik03}. 
A recent progress is its marriage with the idea of 
topology~\cite{Haldane88,Kitaev,ReadGreen,Ivanov,NayakRMP,KaneMele,
Bernevig2006}. A topological many-body state is characterized by 
a bulk quantum number, and the hallmark is the 
existence of an edge state at its interface between a trivial state. 
%Haldane's honeycomb lattice model\cite{Haldane88} plays an important role 
Quantum systems driven by time periodic 
external fields, such as laser, 
is described by "photo-dressed" Floquet states,
and its properties can be different from the original equilibrium 
system~\cite{Oka09,Kitagawa11,Gu11,Lindner11,Dora12,Rudner13,Tenenbaum13,Gedik13,
Rechtsman13,Takayoshi14,Takayoshi13,Jiang11,Liu12,Liu13,Kundu13}.

%Recently a new way of generating quantum nonequilibrium states 
%by using lasers has been intensively 
%discussed~\cite{Oka09,Kitagawa11,Lindner11,Gedik13,Takayoshi14}. 
%In the study, lasers can be viewed as a novel controllable parameter 
%to produce new quantum states. 
%Hamiltonians for quantum systems in external lasers 
%are {\it temporally periodic} and Floquet theorem can be applied to 
%such a periodically driven system. The theorem is the time version of 
%Bloch theorem and it enables us to map a driven system to 
%an effective {\it static} system. After the mapping, one can apply various 
%theoretical techniques for equilibrium states 
%in order to study the driven system. This is a main reason why the study 
%for periodically driven systems has been developed in recent years. 
%Another reason would be that laser science has been 
%rapidly developed and the present laser experiments begin to show 
%a high potential to generate various periodically driven states 
%in a coherent way. 
In a theoretical prediction \cite{Oka09}, it was shown that 
if a circularly polarized laser ({\rm CPL}) is applied to 
two-dimensional Dirac systems, a gap opens at the Dirac point, 
and the system becomes a topological Hall state. 
This gap opening was recently observed by time resolved 
APRES~\cite{Gedik13}. 
%It is important that CPL breaks time reversal symmetry. 
Taking the high frequency limit, it was shown~\cite{Kitagawa11} 
that the effective static model of the 
honeycomb lattice in CPL is equivalent to 
the Haldane honeycomb model for a quantum Hall effect 
without Landau levels~\cite{Haldane88}. 
Chiral edge states induced by CPL were also experimentally observed 
in graphene~\cite{Karch10,Karch11,NoteGanichev} 
as well as in a photonic cousin~\cite{Rechtsman13}.

%%%%%%%%%%%%%%%%%%%%%%%%%%%%%%%%%%%%
\begin{figure}%[tth]
\begin{center}
\includegraphics[width=8cm]{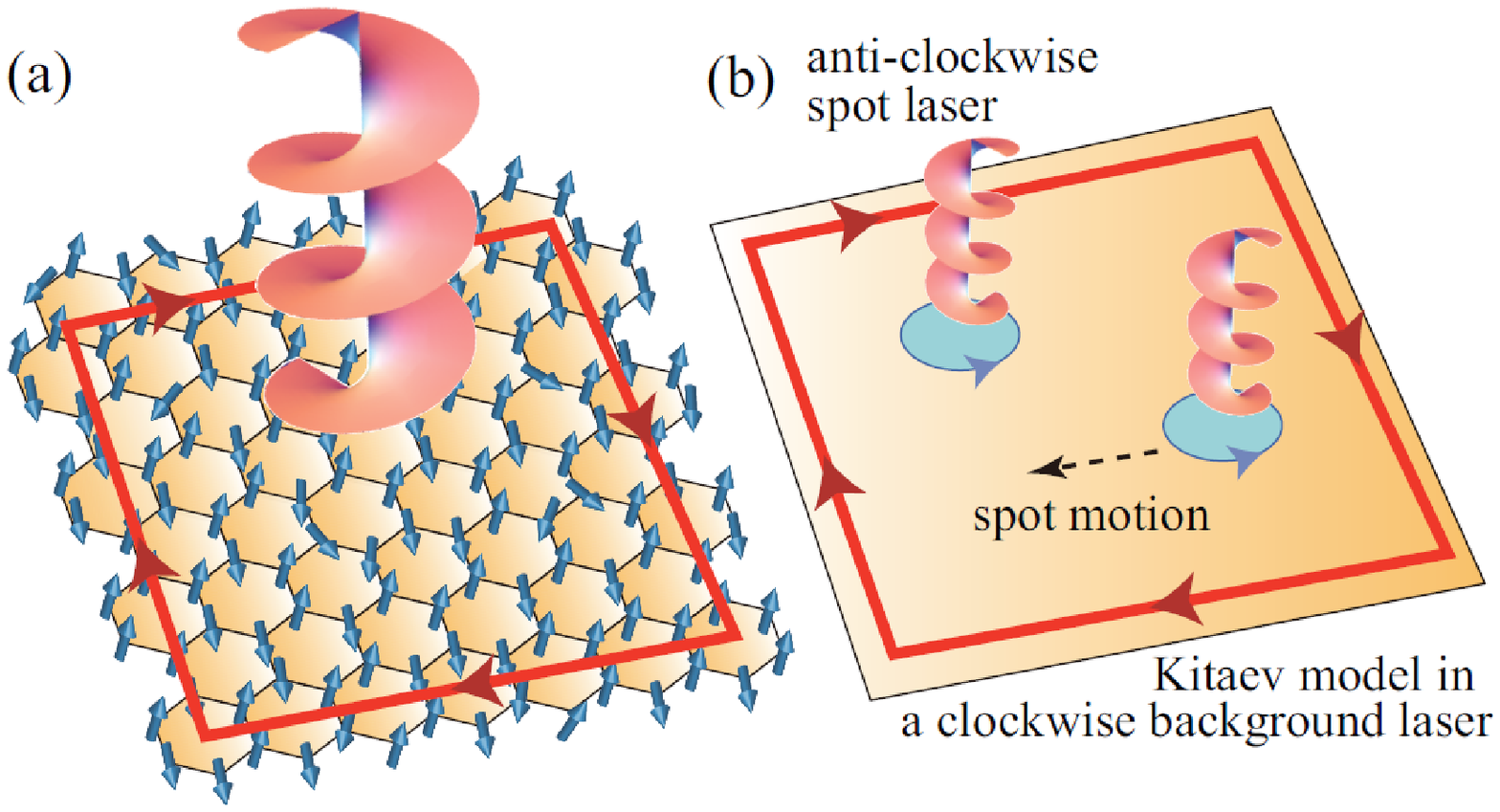}
\end{center}
\caption{(color online) Kitaev model in a circularly or elliptically 
polarized laser. A chiral Majorana edge mode appears. 
(b) Manipulation of edge modes by spatially modulated spot laser.
%(b) Phase diagram of the effective theory for the Kitaev model in a laser. 
%The detail is explained in the text.
}
\label{fig:Kitaev_Laser}
\end{figure}
%%%%%%%%%%%%%%%%%%%%%%%%%%%%%%%%%%%%

In this paper, we propose a quantum spin version of 
the Floquet topological insulator 
in systems described by the Kitaev honeycomb lattice model~\cite{Kitaev}. 
The story is parallel to the electron case~\cite{Oka09,Kitagawa11} 
after fermionizing the spins:
The laser induces an effective {\it imaginary} next nearest neighbor
(NNN) hopping term [see Eq.~(\ref{eq:Laser_Majorana}) later]. 
This plays the same role as the NNN hopping in Haldane model~\cite{Haldane88} 
and opens a topological gap at the Dirac cone of Majorana fermions. 
The Kitaev model is an anisotropic spin-$\frac{1}{2}$ model 
with Ising interactions $S_r^xS_{r'}^x$, $S_r^yS_{r'}^y$ and $S_r^zS_{r'}^z$ 
assigned to the three bonds in the honeycomb lattice. 
Recent studies show that it may be realized as an effective 
low-energy model for a Mott insulator with strong spin-orbit 
coupling~\cite{Jackeli1,Jackeli2}. 
%This model can be fermionized and the ground state and 
%low-energy fermionic excitation spectrum can be exactly calculated. 
In the Kitaev honeycomb lattice model, 
it was shown that a spin-liquid (disordered) ground state is realized 
and fermionic excitations described by a single Dirac cone take place. 
If a gap opens at the Dirac point due to some perturbation, the Kitaev model 
has non-Abelian anyons with an Ising type braiding rule. 
Anyons and their manipulation have attracted much 
attention in the field of quantum computation~\cite{NayakRMP}. 
Kitaev's original proposal for anyons was 
to apply a static magnetic field~\cite{Kitaev}. 
The Zeeman coupling generates a term that opens a gap 
at the third-order perturbation level. 
In this work, we show that it is possible to open a gap 
at the Dirac point using CPL, or more generally, 
elliptically polarized laser (EPL) when a magnetoelectric (ME) coupling 
is taken into account. The virtue of this proposal is twofolds. 
(i) The gap is dynamical, e.g., it can be switched on and off 
as the laser is turned on and off, 
which may lead to possible quantum coherent manipulation. 
(ii) The gap opening term appears at the lowest order. 
This is in contrast to the case of static magnetic fields, 
where a gap appears at the third order and other perturbation terms 
%are difficult to be analyzed and 
may lead to unwanted effects. 
%From the theoretical viewpoint, it is remarkable that the integrability 
%of the system with the laser-induced lowest-order term is maintained. 
%This nature enables us to analytically understand the physics. 
We will show that the integrability of the system with the 
laser-induced lowest-order term is maintained. 
From the theoretical viewpoint, this nature is important to 
analytically understand the laser-driven physics.

The laser-driven gapped state is a spin-liquid version of the 
``Floquet topological insulator"~\cite{Oka09,Kitagawa11,Gu11,Lindner11}. 
In the state, a chiral edge mode of Majorana fermions~\cite{ReadGreen} 
appears as indicated in Fig.~\ref{fig:Kitaev_Laser}(a). 
The direction of this mode can be switched by altering the helicity of the laser. 
It is also possible to make islands of topological spin-liquid states 
using a spatial modulation technique~\cite{Gedik03,Tenenbaum13}
[see Fig.~\ref{fig:Kitaev_Laser}(b)]: 
A spot of EPL leads to a locally topological state in the "sea" 
of the Kitaev model. The topological island can be spatially manipulated 
by slowly moving the spot position.

%Coupling between magnetic exchange interaction and 
%external electirc field is generally expected to exist in any magnetic 
%materials (although its coupling constant would depend strongly on 
%their microscopic detail). 
%We will show that for the Kitaev model with 
%this type of ME coupling, an elliptically or circularly polarized laser 
%can induce a nonequilibrium gapped spin-liquid state with a gapless 
%chiral edge mode and the direction of the edge current can be 
%switched by inverting the rotating direction of the laser. 
%We will show that the effective Hamiltonian for he Kitaev model with 
%this ME coupling in an elliptically polarized laser is exactly solvable, 
%and it leads to a gapped spin-liquid state with a gapless 
%chiral edge mode. The edge-current direction can be 
%switched by inverting the laser-rotating direction. 

%%%%%%%%%%%%%%%%%%%%%%%%%%%%%%%%%%%%%%%%%%%%%%%%%%%%%%%%%%%%
%%%%%%%%%%%%%%%%%%%%%%%%%%%%%%%%%%%%%%%%%%%%%%%%%%%%%%%%%%%%
%%%%%%%%%%%%%%%%%%%%%%%%%%%%%%%%%%%%%%%%%%%%%%%%%%%%%%%%%%%%

%%%%%%%%%%%%%%%%%%%%%%%%%%%%%%%%%%%%
\begin{figure}%[tth]
\begin{center}
\includegraphics[width=7cm]{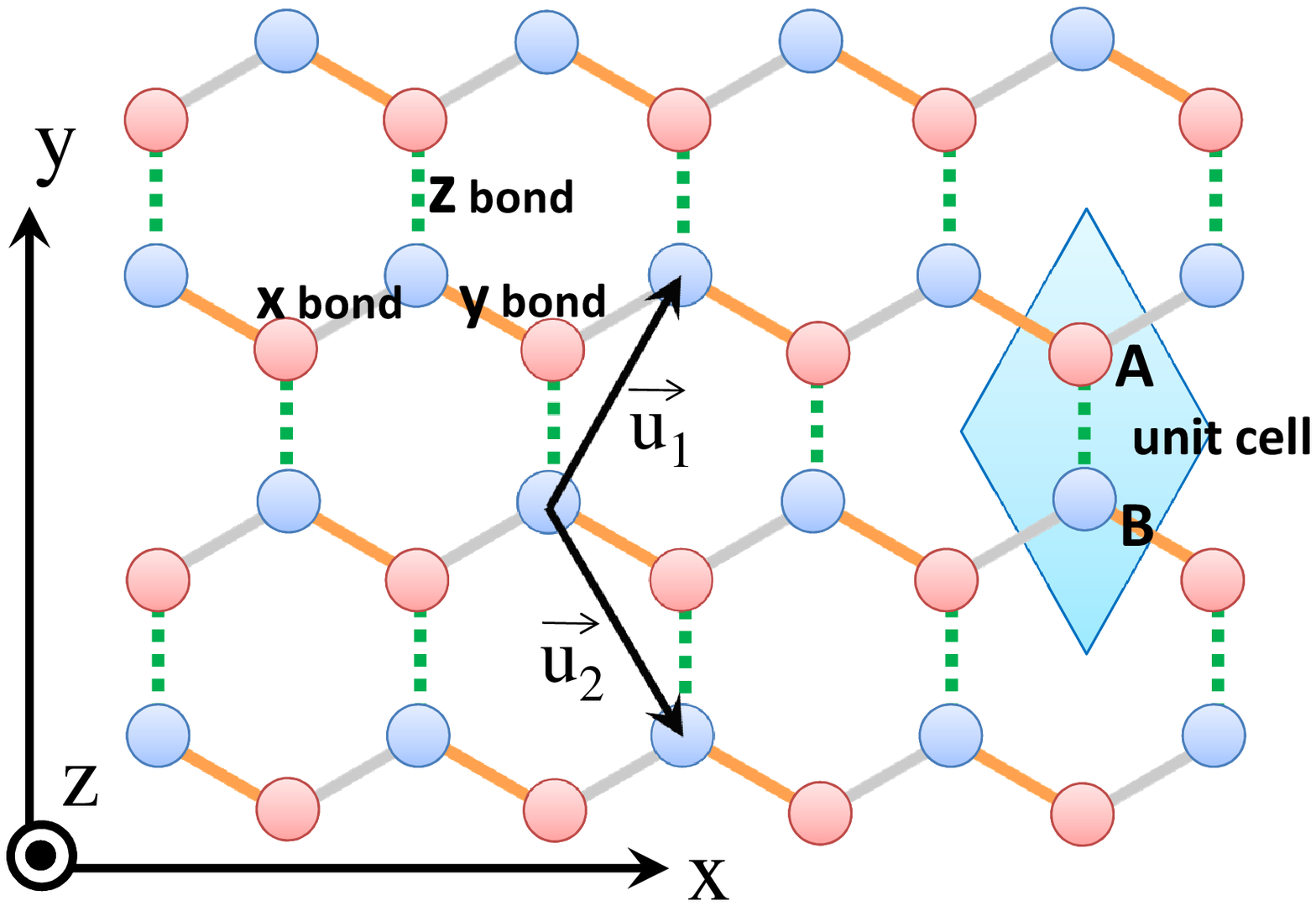}
\end{center}
\caption{(color online) Honeycomb lattice of the Kitaev model. 
There are three types of bonds ($x$, $y$, and $z$ bonds). 
The unit vectors are defined by ${\bol u}_1=\hat x/2 +\sqrt{3}\hat y/2$ and 
${\bol u}_2=\hat x/2 -\sqrt{3}\hat y/2$, where $\hat x=(1,0)$ and 
$\hat y=(0,1)$ ($|{\bol u}_{1,2}|=1$). 
The unit cell consists of two neighboring sites 
and the system is divided into A and B sublattices.}
\label{fig:KitaevLattice}
\end{figure}
%%%%%%%%%%%%%%%%%%%%%%%%%%%%%%%%%%%%

%\section{Modeling}
{\it Model $-$}
We focus on the Kitaev model with a ME coupling in EPL 
(see Fig.~\ref{fig:Kitaev_Laser}). 
We assume that the ME coupling originates from 
electric or phonon-assisted polarization 
on each bond and its strength is proportional to the exchange 
interaction of the bond. 
Then, a quantum spin system responds to an external 
{\it electric} field through the polarization tensor. 
%in addition to a magnetic field. 
This effect is known as magneto striction and has been 
confirmed in various multiferroic 
materials~\cite{Wang,Pimenov06,Miyahara,Mochizuki,Furukawa}.
Our total Hamiltonian is given by
\begin{eqnarray}
\hat {\cal H}(t) &=& \hat {\cal H}_{\rm Kitaev}
+\hat {\cal H}_{\rm ME}(t).%+\hat {\cal H}_{\rm M}(t).
\label{eq:Kitaev_inLaser}
\end{eqnarray}
The first term is the Kitaev Hamiltonian 
\begin{eqnarray}
\hat {\cal H}_{\rm Kitaev} &=& \sum_{\alpha=x,y,z}
J_{\alpha}\sum_{\langle {\bol r},{\bol r'}\rangle_\alpha}
\sigma_{\bol r}^\alpha \sigma_{\bol r'}^\alpha.
\label{eq:Kitaev}
\end{eqnarray}
This model is defined on a honeycomb lattice as displayed in 
Fig.~\ref{fig:KitaevLattice}. 
Here $S_{\bol r}^\alpha=\frac{1}{2}\sigma_{\bol r}^\alpha$ 
denotes the $\alpha$ component of spin-$\frac{1}{2}$ operator 
on site ${\bol r}$ ($\sigma_{\bol r}^{x,y,z}$ are Pauli matrices and 
we set $\hbar=1$), $\sum_{\langle {\bol r},{\bol r'}\rangle_\alpha}$ 
stands for the summation over all neighboring spin pairs on 
$\alpha$ bonds (see Fig.~\ref{fig:KitaevLattice}), 
and $J_\alpha$ is the Ising coupling constant of the $\alpha$ bond. 
%The honeycomb lattice, its coordinate, unit vectors 
%${\bol u}_{1,2}$, and $x$, $y$, and $z$ bonds are all defined 
%in Fig.~\ref{fig:KitaevLattice}. 
In order to describe the ME coupling, we introduce 
the electric polarization ${\bol P}_{({\bol r},{\bol r'})_\alpha}$ 
on each $\alpha$ bond $({\bol r},{\bol r'})_\alpha$ %that is
proportional to the Ising interaction
\begin{eqnarray}
{\bol P}_{({\bol r},{\bol r'})_\alpha}&=& {\bol \pi}_\alpha 
\sigma_{\bol r}^\alpha \sigma_{\bol r'}^\alpha,
\label{eq:Polarization}
\end{eqnarray}
where ${\bol \pi}_\alpha$ is the ME-coupling vector. % and its strength usually depends on laser frequency. 
%where the vector 
%${\bol \pi}_\alpha $ depend on only $\alpha$ and does not on 
%the bond position $({\bol r},{\bol r'})$. 
Using the total polarization ${\bol P}_{\rm tot}=
\sum_\alpha{\bol P}_{{\rm tot},\alpha}
=\sum_\alpha\sum_{\langle {\bol r},{\bol r'}\rangle_\alpha}
{\bol P}_{({\bol r},{\bol r'})_\alpha}$,
the ME term becomes
\begin{eqnarray}
\hat {\cal H}_{\rm ME}(t)=- {\bol E}(t)\cdot{\bol P}_{\rm tot}.
\end{eqnarray}
We assume that the electric field vector ${\bol E}$ of 
the EPL is in the $x$-$y$ plane (= plane of the honeycomb lattice) and given by \begin{eqnarray}
{\bol E}(t) &=& %A_0\Omega 
E(\mp\cos(\Omega t+\delta),\sin(\Omega t),0),
%\nonumber\\
%{\bol B}(t) &=& A_0\Omega c^{-1}(-\sin(\Omega t),\mp\cos(\Omega t+\delta),0),
\label{eq:LaserFields}
\end{eqnarray}
where $t$ is time, $\Omega$ is the laser frequency, $E$ is the magnitude 
of field and 
%$c$ is light speed and 
the signs $\mp$ denote clockwise/anticlockwise rotating laser. 
The phase $\delta$ ($|\delta|\leq \pi/2$) controls the ellipticity; 
$\delta=0$ and $\pi/2$ correspond to circularly and linearly 
polarized laser, respectively. 
%($\hat {\cal H}_{\rm M}(t)=g{\bol B}(t)\cdot{\bol S}_{\rm tot}$) 
%($g$ is the effective gyromagnetic tensor). 
We note that laser has a magnetic field component 
that couples to spin systems via Zeeman interaction. 
However, we can ignore this because the
magnetic fields are much weaker than the electric fields
in electromagnetic waves. 
%This is a possible condition for real magnetic materials. 
%We will show that a topological spin liquid state appears in the model. 
%Hereafter we assume that the ME coupling constant 
%$A_0\Omega\epsilon|{\bol \pi}_\alpha|$ is much larger than 
%the Zeeman coupling one $A_0\omega c^{-1}g$, and neglect 
%the Zeeman term $\hat {\cal H}_{\rm M}(t)$. 

%%%%%%%%%%%%%%%%%%%%%%%%%%%%%%%%%%%%%%%%%%%%%%%%%%%%%%%%%%%%
%%%%%%%%%%%%%%%%%%%%%%%%%%%%%%%%%%%%%%%%%%%%%%%%%%%%%%%%%%%%
%%%%%%%%%%%%%%%%%%%%%%%%%%%%%%%%%%%%%%%%%%%%%%%%%%%%%%%%%%%%
{\it Floquet theory and $1/\Omega$ expansion $-$}
The Hamiltonian of Eq.~(\ref{eq:Kitaev_inLaser}) is temporally periodic: 
$\hat {\cal H}(t+T)=\hat {\cal H}(t)$ with $T=2\pi/\Omega$. 
%Let us apply Floquet theorem to Eq.~(\ref{eq:Kitaev_inLaser}). 
The Floquet theorem, i.e., temporal version of the Bloch theorem,
states that the time-dependent 
Schr\"odinger equation $i\partial_t|\Psi(t)\rangle 
=\hat {\cal H}(t)|\Psi(t)\rangle$ can be mapped to an 
effective {\it static} eigenvalue problem 
$\sum_n(\hat {\cal H}_{m-n}-m\Omega\delta_{m,n})|\Phi^n\rangle 
= \epsilon|\Phi^m\rangle$,
which can be proved by Fourier transform
$|\Phi^m\rangle=T^{-1}\int dt e^{im\Omega t}|\Phi(t)\rangle$ 
and $\hat{\cal H}_m=T^{-1}\int dt e^{im\Omega t}\hat {\cal H}(t)$. 
The time periodic 
Floquet state $|\Phi(t)\rangle(=|\Phi(t+T)\rangle)$ is related to the 
solution of the time-dependent Schr\"odinger equation 
via $|\Psi(t)\rangle=e^{-i\epsilon t}|\Phi(t)\rangle$
($\epsilon$ is called the Floquet quasi-energy). 
%Floquet theorem says that the time-dependent Schr\"odinger 
%equation is mapped into an effective {\it static} 
%eigenvalue problem,
%\begin{eqnarray}
%\sum_n(\hat {\cal H}_{m-n}-m\Omega\delta_{m,n})|\Phi^n\rangle &=& 
%\epsilon|\Phi^m\rangle
%\label{eq:Floquet}
%\end{eqnarray}
In the present case, we have 
$\hat{\cal H}_0=\hat{\cal H}_{\rm Kitaev}$, 
$\hat{\cal H}_{+1}=-\frac{E}{2}(\mp e^{-i\delta},i,0)\cdot{\bol P}_{\rm tot}$, 
and $\hat{\cal H}_{-1}=-\frac{E}{2}(\mp e^{i\delta},-i,0)
\cdot{\bol P}_{\rm tot}$. 
%$A=\tilde \epsilon A_0\Omega$. 
Terms $\hat{\cal H}_{\pm p}$ with $p\geq 2$ are zero.

\begin{figure}%[tth]
\begin{center}
\includegraphics[width=8cm]{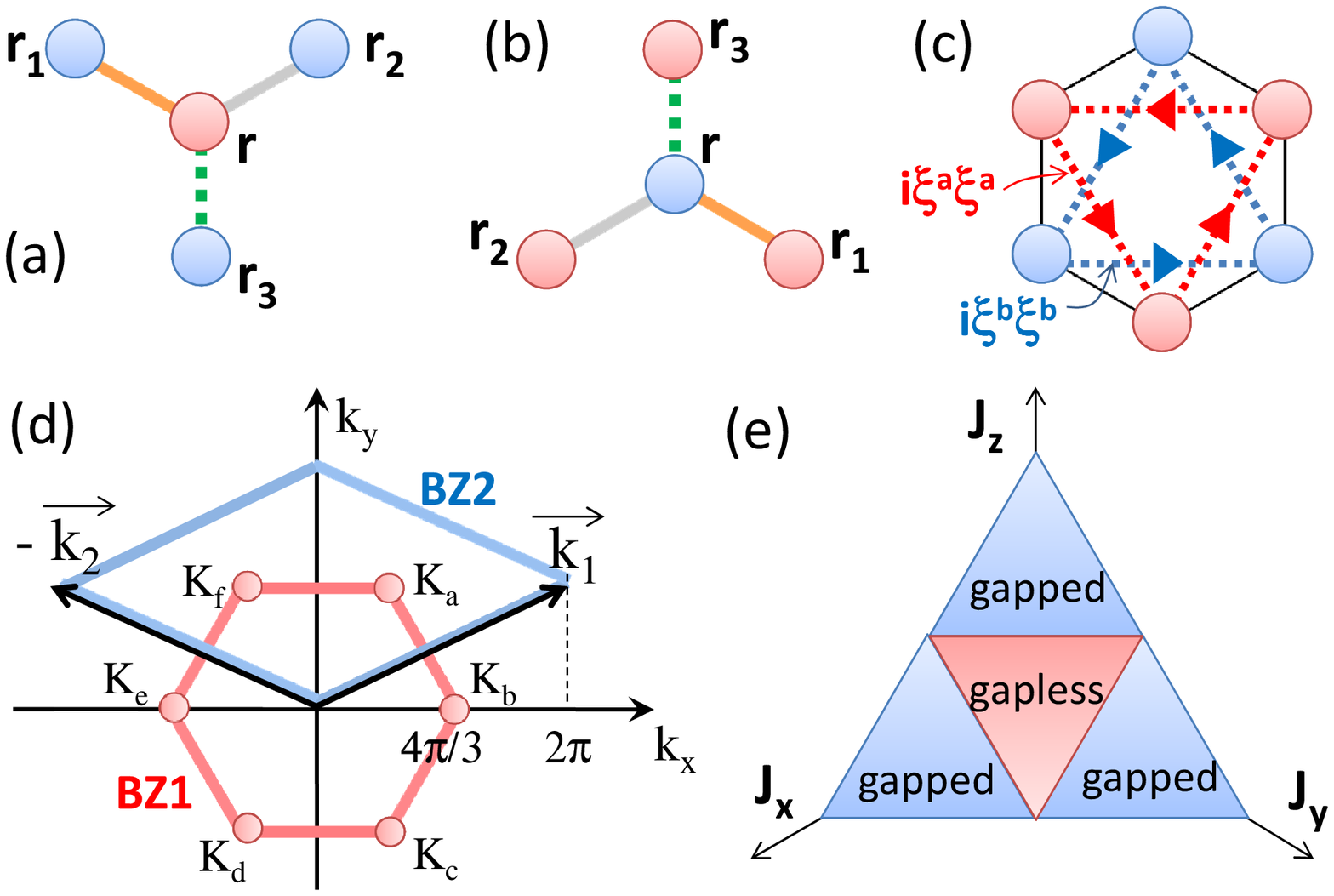}
\end{center}
\caption{(color online) Three sites around site ${\bol r}\in A$ (a)
and around ${\bol r}\in B$ (b). 
(c) NNN coupling in Eq.~(\ref{eq:Laser_Majorana}). 
(d) Brillouin zones (BZ) of the Kitaev model, hexagonal (BZ1) 
and rhombus (BZ2), where ${\bol k}_{1,2}$ are reciprocal lattice vectors. 
(e) Ground-state phase diagram of $\hat{\cal H}_{\rm Kitaev}$ 
in the plane $J_x+J_y+J_z=1$~\cite{Kitaev,Chen}. 
}
\label{fig:Hexagon}
\end{figure}

We consider the case where $\Omega$ is much larger than the 
energy scale of $\hat{\cal H}_{\rm Kitaev}$ (i.e., $|J_{x,y,z}|$). 
Since the typical energy of spin couplings is in the THz 
regime (1THz$\sim$4meV), this implies that THz or mid-infrared lasers 
are suitable. Then, each $m$-photon subspace 
are energetically isolated from other subspaces 
and the off-diagonal terms 
$\hat{\cal H}_{\pm 1}$ can be treated perturbatively.
The effective Hamiltonian acting on the $0$-photon subspace
is given by~\cite{Kitagawa11} 
\begin{eqnarray}
\hat{\cal H}_{\rm eff} &=&  \hat{\cal H}_{\rm Kitaev} 
-\frac{1}{\Omega}[\hat{\cal H}_{+1},\hat{\cal H}_{-1}]
%+{\cal O}(\Omega^{-2}),
\label{eq:Effective}
\end{eqnarray}
up to ${\cal O}(\Omega^{-1})$. 
Utilizing Eq.~(\ref{eq:Polarization}), we can show that the first order term 
$\hat{\cal H}_{\Omega}=-[\hat{\cal H}_{+1},\hat{\cal H}_{-1}]/\Omega$ results 
in the following three-spin interactions:
\begin{eqnarray}
\hat{\cal H}_{\Omega} &=& \pm \frac{1}{\Omega}E^2 \cos\delta \Big[
G_{12}\Big(\sum_{{\bol r}\in A}
\sigma_{{\bol r}_1}^x\sigma_{{\bol r}}^z\sigma_{{\bol r}_2}^y
+\sum_{{\bol r}\in B}
\sigma_{{\bol r}_1}^x\sigma_{{\bol r}}^z\sigma_{{\bol r}_2}^y\Big)
\nonumber\\
&&+G_{23}\Big(\sum_{{\bol r}\in A}
\sigma_{{\bol r}_2}^y\sigma_{{\bol r}}^x\sigma_{{\bol r}_3}^z
+\sum_{{\bol r}\in B}
\sigma_{{\bol r}_2}^y\sigma_{{\bol r}}^x\sigma_{{\bol r}_3}^z\Big)
\nonumber\\
&&+G_{31}\Big(\sum_{{\bol r}\in A}
\sigma_{{\bol r}_3}^z\sigma_{{\bol r}}^y\sigma_{{\bol r}_1}^x
+\sum_{{\bol r}\in B}
\sigma_{{\bol r}_3}^z\sigma_{{\bol r}}^y\sigma_{{\bol r}_1}^x\Big)
\Big].
\label{eq:Commutator}
\end{eqnarray}
Here, $G_{\alpha\beta}=\hat z\cdot({\bol \pi_\alpha
\times{\bol \pi}_\beta})$ [$\hat z=(0,0,1)$ and symbol $\times$ denotes 
outer product], $\sum_{{\bol r}\in A(B)}$ summation over 
sublattice A (B), and the signs $\mp$ corresponds to 
clockwise/anticlockwise rotating laser respectively. 
Vectors ${\bol r}_{1,2,3}$ are sites around ${\bol r}$ as
depicted in Fig.~\ref{fig:Hexagon}. 
%Equation~(\ref{eq:Commutator}) indicates that a laser adds three spin 
%interactions to the Kitaev model as $\Omega$ is sufficiently large. 
Note that $\hat{\cal H}_{\Omega}$ is non-zero unless the laser is 
linearly polarized ($\delta=\pi/2$) or all $\bol \pi_{x,y,z}$ 
are parallel with each other. 
This laser-induced three-spin interaction (\ref{eq:Commutator})
is the main result of this paper 
and we will discuss its physical outcomes in the following.
% show that it generates a topological state in the Kitaev model. 

%%%%%%%%%%%%%%%%%%%%%%%%%%%%%%%%%%%%
%%%%%%%%%%%%%%%%%%%%%%%%%%%%%%%%%%%%

%%%%%%%%%%%%%%%%%%%%%%%%%%%%%%%%%%%%%%%%%%%%%%%%%%%%%%%%%%%%
%%%%%%%%%%%%%%%%%%%%%%%%%%%%%%%%%%%%%%%%%%%%%%%%%%%%%%%%%%%%
%%%%%%%%%%%%%%%%%%%%%%%%%%%%%%%%%%%%%%%%%%%%%%%%%%%%%%%%%%%%
{\it Fermionization and Floquet topological state $-$} 
Here, we study the low-energy effective model (\ref{eq:Effective}) 
by fermionization~\cite{Kitaev,Chen}. 
Jordan-Wigner transformation maps spin-$\frac{1}{2}$ operators 
to fermions:
\begin{eqnarray}
2c_{\bol r}^\dag c_{\bol r}-1 = \sigma^z_{\bol r},&&
%\nonumber\\
c_{\bol r}^\dag = \frac{1}{2}
\Big[\prod_{{\bol r}'<{\bol r}}\sigma_{{\bol r}'}^z\Big]
\sigma^+_{\bol r}.
\label{eq:JW}
\end{eqnarray}
Here $\sigma_{\bol r}^\pm=\sigma_{\bol r}^x\pm i\sigma_{\bol r}^y$ 
and $\prod_{{\bol r}'<{\bol r}}\sigma_{{\bol r}'}^z$ is a 
non-local product whose path is defined in Fig. 2 of Ref.~\onlinecite{Chen}. 
Operators $c_{\bol r}$ and $c_{\bol r}^\dag$ are 
complex fermions satisfying $\{c_{\bol r},c_{{\bol r}'}^\dag\}
=\delta_{{\bol r},{\bol r}'}$ and 
$\{c_{\bol r},c_{{\bol r}'}\}
=\{c_{\bol r}^\dag,c_{{\bol r}'}^\dag\}=0$. 
We define four Majorana (real) fermions
\begin{eqnarray}
\xi_{\bol r}^a=-i(a_{\bol r}-a_{\bol r}^\dag), 
&& \chi_{\bol r}^a=a_{\bol r}+a_{\bol r}^\dag,
\nonumber\\ 
\chi_{\bol r}^b=-i(b_{\bol r}+b_{\bol r}^\dag),&&
\xi_{\bol r}^b=b_{\bol r}+b_{\bol r}^\dag,
\label{eq:Majorana}
\end{eqnarray}
where $a_{\bol r}$ ($b_{\bol r}$) is a fermion 
$c_{\bol r}$ on sublattice A (B). 
We choose $\xi_{\bol r}^a$ and $\xi_{\bol r}^b$ 
($\chi_{\bol r}^a$ and $\chi_{\bol r}^b$) to be in the same unit cell. 
Majorana fermions satisfy 
$\{\xi_{\bol r}^\alpha,\xi_{\bol r'}^\beta\}
=\{\chi_{\bol r}^\alpha,\chi_{\bol r'}^\beta\}
=2\delta_{\alpha,\beta}\delta_{\bol r,\bol r'}$, 
$\{\xi_{\bol r}^\alpha,\chi_{\bol r'}^\beta\}=0$, 
$\xi^\dag=\xi$ and $\chi^\dag=\chi$. 
Using these fermions, the Kitaev Hamiltonian $\hat{\cal H}_{\rm Kitaev}$
becomes~\cite{note1} 
\begin{eqnarray}
\hat{\cal H}_{\rm Kitaev} = 
\sum_{\bol r}iJ_x\xi_{\bol r}^a\xi_{\bol r +\bol u_1}^b
-iJ_y\xi_{\bol r}^b\xi_{\bol r +\bol u_2}^a
-\hat I_{\bol r}iJ_z\xi_{\bol r}^b\xi_{\bol r}^a,
%\nonumber\\
\label{eq:Kitaev_Majorana}
\end{eqnarray}
where $\bol r=n_1\bol u_1+n_2\bol u_2$ ($n_{1,2}$: integer), 
$\sum_{\bol r}$ runs over the unit cells. 
The $\chi$ fermion appears in the Hamiltonian through a
locally conserved operator
$\hat I_{\bol r}=i\chi_{\bol r}^a\chi_{\bol r}^b$ 
with an eigenvalue $\pm 1$. 
In the ground state of $\hat{\cal H}_{\rm Kitaev}$, 
it is known~\cite{Kitaev,Chen,Lieb} that 
we can set all $\hat I_{\bol r}$ to be 
unity (or $-1$) and vortex-type excitations defined by 
sign flips of $\hat I_{\bol r}$ are gapped 
(i.e., fermions $\chi^{a,b}$ are gapped). We will set 
$\hat I_{\bol r}=1$ hereafter.

The laser-induced three-spin interaction (\ref{eq:Commutator})
can be fermionized as well and results in
\begin{eqnarray}
\hat{\cal H}_{\Omega} &=& \pm \frac{1}{\Omega}E^2
\sum_{\bol r} iG_{12}\Big[\xi_{\bol r}^b \xi_{\bol r+\bol u_1+\bol u_2}^b 
+\xi_{\bol r}^a \xi_{\bol r-\bol u_1-\bol u_2}^a\Big]
\nonumber\\
&&+i\hat I_{\bol r}G_{23}\Big[\xi_{\bol r}^b \xi_{\bol r-\bol u_2}^b
+\xi_{\bol r}^a \xi_{\bol r+\bol u_2}^a\Big]
\nonumber\\
&&+i\hat I_{\bol r}G_{31}\Big[\xi_{\bol r}^b \xi_{\bol r-\bol u_1}^b
+\xi_{\bol r}^a \xi_{\bol r+\bol u_1}^a\Big].
\label{eq:Laser_Majorana}
\end{eqnarray}
The Kitaev Hamiltonian (\ref{eq:Kitaev_Majorana}) 
represents nearest neighbor hopping, while 
$\hat{\cal H}_{\Omega}$ describes an imaginary NNN hopping 
as displayed in Fig.~\ref{fig:Hexagon}(c)~\cite{note2}.

\begin{figure}%[tth]
\begin{center}
\includegraphics[width=8cm]{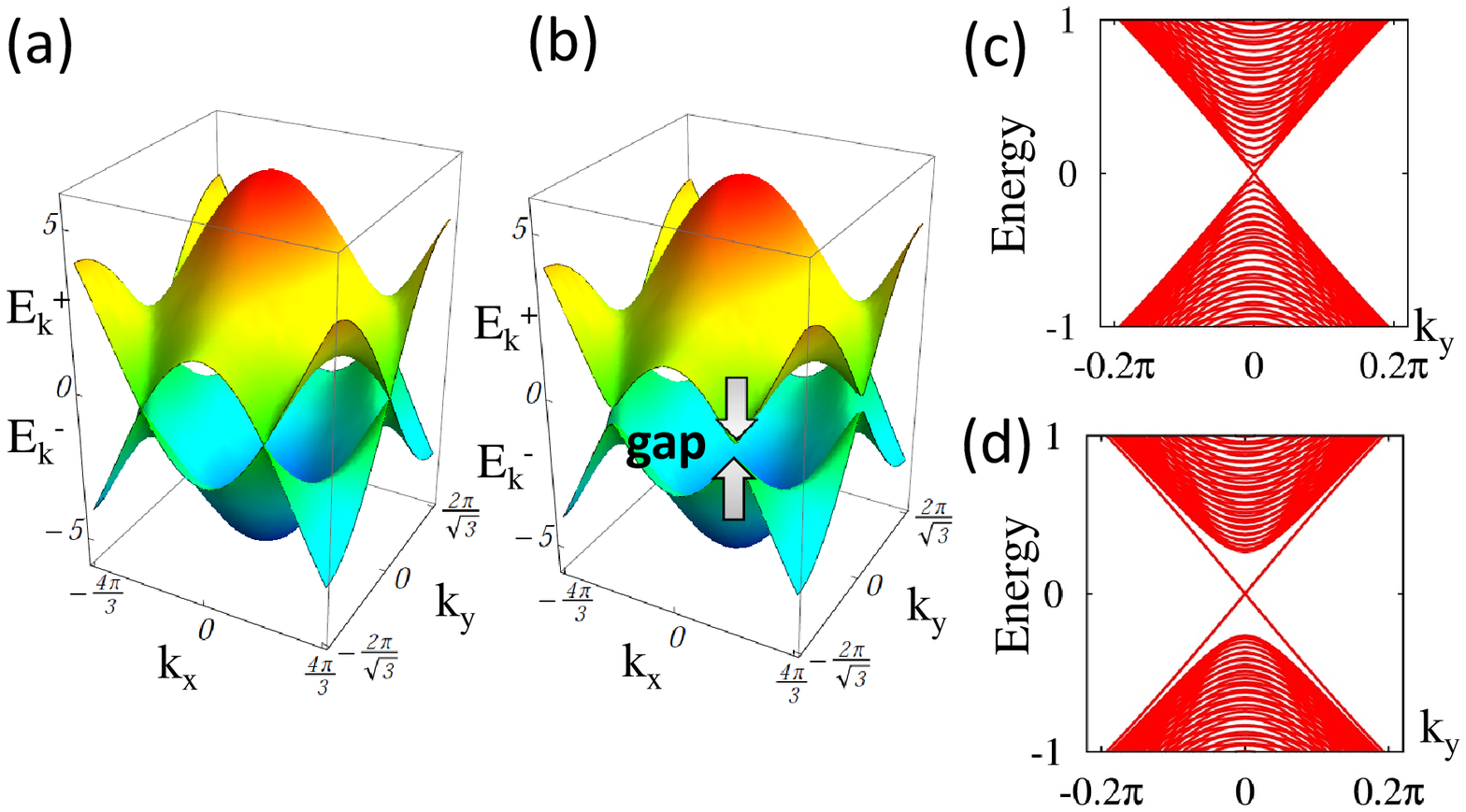}
\end{center}
\caption{(color online) (a)(b) Fermion dispersions 
of the effective model (\ref{eq:Effective}) 
in the  $\hat I_{\bol r}=1$ sector under,
and (c)(d) is the spectrum in cylindrical geometry with an armchair 
edge along the $y$ direction. Panels (a) and (c) [(b) and (d)] are 
results for the zero electric field case ($E=0$) [a finite-field case 
($4E^2\Omega^{-1}\cos\delta=0.1$)]. 
Two gapless modes connecting the bulk continuum in panel (d) 
correspond to chiral edge modes along the $y$ direction.
$J_{x,y,z}=1$ and $G_{12,23,31}=1$ 
are used.}
\label{fig:Dispersion}
\end{figure}

To see the spectral nature in the $\hat I_{\bol r}=1$ sector, 
we move to the momentum $\bol k$ space using 
Fourier transforms $\xi^{a(b)}_{\bol r}
=\sqrt{2/N}\sum'_{\bol k}[e^{\bol k\cdot\bol r}\xi^{a(b)}_{\bol k}
+e^{-\bol k\cdot\bol r}\xi^{a(b)\,\,\dag}_{\bol k}]$. 
Here $\sum'_{\bol k}$ stands for summation over {\it half} 
the Brillouin zone (BZ) and $N$ is the total number of unit cells.
We can choose either hexagonal or rhombus form as the BZ 
shown in Fig.~\ref{fig:Hexagon}(d). Fermions in the 
$\bol k$ space are of complex type, and satisfy 
$\{\xi^\alpha_{\bol k},\xi^{\beta\dag}_{\bol k'}\}=
\delta_{\alpha,\beta}\delta_{\bol k,\bol k'}$ and 
$\{\xi^\alpha_{\bol k},\xi^{\beta}_{\bol k'}\}=0$. 
We can denote 
$\hat{\cal H}_{\rm eff} = \sum'_{\bol k}
\Xi^\dag_{\bol k} {\cal H}_{\bol k}\Xi_{\bol k}$, 
%\begin{eqnarray}
%\hat{\cal H}_{\rm Kitaev} &=& \sum_{\bol k}{}^{'}
%\Xi^\dag_{\bol k} {\cal H}_{\bol k}\Xi_{\bol k}, 
%\label{eq:Hamiltonian_kspace}
%\end{eqnarray}
where $\Xi_{\bol k}={}^t(\xi^a_{\bol k},\xi^b_{\bol k})$ and 
the $2\times 2$ matrix ${\cal H}_{\bol k}$ is given by
\begin{eqnarray}
{\cal H}_{\bol k}&=& \sum_{\alpha=x,y,z}h_{\bol k}^\alpha \tau^\alpha.
\label{eq:Hamiltonian_matrix}
\end{eqnarray}
Here $\tau^{x,y,z}$ are Pauli matrices, 
$h_{\bol k}^x=-2(J_x\sin(\bol k\cdot\bol u_1)-J_y\sin(\bol k\cdot\bol u_2))$, 
$h_{\bol k}^y=-2(J_x\cos(\bol k\cdot\bol u_1)
+J_y\cos(\bol k\cdot\bol u_2)+J_z)$ and 
$h_{\bol k}^z=\pm\frac{4}{\Omega}E^2\cos\delta
[G_{12}\sin(\bol k\cdot(\bol u_1+\bol u_2))-G_{23}\sin(\bol k\cdot\bol u_2)
-G_{31}\sin(\bol k\cdot\bol u_1)]$. 
We stress that EPL gives a non-zero $z$-component $h_{\bol k}^z$. 
The Hamiltonian~(\ref{eq:Hamiltonian_matrix}) can be diagonalized 
leading to dispersions 
%\begin{eqnarray}
%E_{\bol k}^{\pm}&=& \pm(|h_{\bol k}^x|+|h_{\bol k}^y|+|h_{\bol k}^z|)
%\label{eq:dispersion}
%\end{eqnarray}
\begin{eqnarray}
E_{\bol k}^{\pm}&=& 
\pm(|h_{\bol k}^x|^2+|h_{\bol k}^y|^2+|h_{\bol k}^z|^2)^{1/2}.
\label{eq:Dispersion_Topo}
\end{eqnarray}
We plot the energy spectrum for the zero field case ($E=0$) at 
$J_x=J_y=J_z$ in Fig.~\ref{fig:Dispersion}(a). 
We see gapless Dirac cones appearing at the six corners 
$K_{\rm a,\cdots,f}$ of BZ1. However, since we defined 
two complex fermions in momentum space from two real ones 
$\xi^{a,b}_{\bol r}$, there is a redundancy 
and we should restrict ourselves to half 
the BZ [e.g., the right triangular area of BZ2 in 
Fig.~\ref{fig:Hexagon}(d)]. 
Thus, only gapless excitations around a single Dirac point 
(e.g., $K_{\rm a}$) describes the low-energy physics of the Kitaev model. 
The gapless Dirac cone exists
if the condition $|J_\alpha|\leq |J_\beta|+|J_\gamma|$~\cite{Kitaev}
%\begin{eqnarray}
%|J_\alpha|\leq |J_\beta|+|J_\gamma|
%\label{eq:gapless}
%\end{eqnarray}
is satisfied ($\alpha$, $\beta$ and $\gamma$ are all different). 
The ground-state phase diagram of the Kitaev model is summarized 
as in Fig.~\ref{fig:Hexagon}(e). In the gapped phase, 
vortex excitations of $\chi^{a,b}$ are regarded 
as Abelian anyons~\cite{Kitaev}.

%%%%%%%%%%%%%%%%%%%%%%%%%%%%%%%%%%%%
%%%%%%%%%%%%%%%%%%%%%%%%%%%%%%%%%%%%

%%%%%%%%%%%%%%%%%%%%%%%%%%%%%%%%%%%%%%%%%%%%%%%%%%%%%%%%%%%%
%%%%%%%%%%%%%%%%%%%%%%%%%%%%%%%%%%%%%%%%%%%%%%%%%%%%%%%%%%%%
%%%%%%%%%%%%%%%%%%%%%%%%%%%%%%%%%%%%%%%%%%%%%%%%%%%%%%%%%%%%
%{\it Laser-induced topological state} $-$ 

Next, we move to the finite field case $E\ne 0$. 
We show that {\it $\hat{\cal H}_{\Omega}$ 
plays the role similar to the NNN hopping in the Haldane honeycomb 
model}~\cite{Haldane88}. 
The energy dispersion is plotted in Fig.~\ref{fig:Dispersion}(b). 
The laser-induced NNN hopping opens a finite gap at the Dirac point
in the gapless phase. Let us investigate the topological nature of 
the laser-driven gapped state. To this end, 
we first focus on the low-energy physics around the Dirac point 
$K_{\rm a}$. The Hamiltonian matrix near 
${\bol k}_a=(2\pi/3,2\pi/\sqrt{3})$ is given by ($J_{x,y,z}=J$)
\begin{eqnarray}
{\cal H}_{\bol k_a+\delta \bol k} = 
\left(
\begin{array}{cc}
 m & \sqrt{3}J( i\delta k_x+\delta k_y)\\
\sqrt{3}J(-i\delta k_x+\delta k_y) & -m
\end{array}
\right)
%\hspace{1cm}
\label{eq:DiracHamiltonian}
\end{eqnarray}
up to linear order in $\delta {\bol k}={\bol k}-{\bol k}_a$. 
The mass parameter is 
$m=\pm 2\sqrt{3}E^2\Omega^{-1}\cos\delta (G_{12}+G_{23}+G_{31})$ and 
its sign $\pm$ corresponds to clockwise/anticlockwise rotation of laser.
This is nothing but a Hamiltonian for a $(2+1)$-dimensional Dirac 
fermion with a gap $2m$. The mass $m$ is generally non-zero except for 
$\delta=\pi/2$ or $G_{12}+G_{23}+G_{31}=0$. 
It is known that a massive Dirac fermion exhibits an 
anomalous quantized "Hall" effect without Landau levels~\cite{Haldane88}. 
Therefore, Eq.~(\ref{eq:DiracHamiltonian}) indicates that 
a gapless chiral edge mode of Majorana fermions is induced by EPL and
its direction can be changed by inverting the helicity of the laser. 
We stress that the present edge mode is chargeless in contrast to 
the case of integer quantum Hall effects (IQHE). 
%since Majorana fermions do not have any charge. 
In Figs.~\ref{fig:Dispersion} (c) and (d), 
we explicitly show the energy levels of the Kitaev model in EPL 
defined on a cylinder geometry with an armchair edge along the $y$ direction. 
When the field is turned on, we see the gap opening 
as well as a formation of the edge mode. 
We confirmed that the edge mode 
is stable against small change of $J_{x,y,z}$ and $E$. 
This edge mode is a Majorana type. This can be confirmed 
by mapping $\hat{\cal H}_{\rm eff}$ to a Bogoliubov-de-Gennes (BdG) 
Hamiltonian for topological superconductor through new fermions 
$d_{\bol r}=(\xi^a_{\bol r}+i\xi^b_{\bol r})/2$~\cite{Chen}. 
The dispersion of the BdG Hamiltonian on a cylinder geometry 
has a gapless edge mode, which clearly shows the presence of 
a Majorana edge mode~\cite{ReadGreen}. 
It is known~\cite{Kitaev} that the laser-induced topological phase 
has gapped non-Abelian anyon excitations originating 
from fermions $\chi^{a,b}$.

\begin{figure}%[tth]
\begin{center}
\includegraphics[width=8cm]{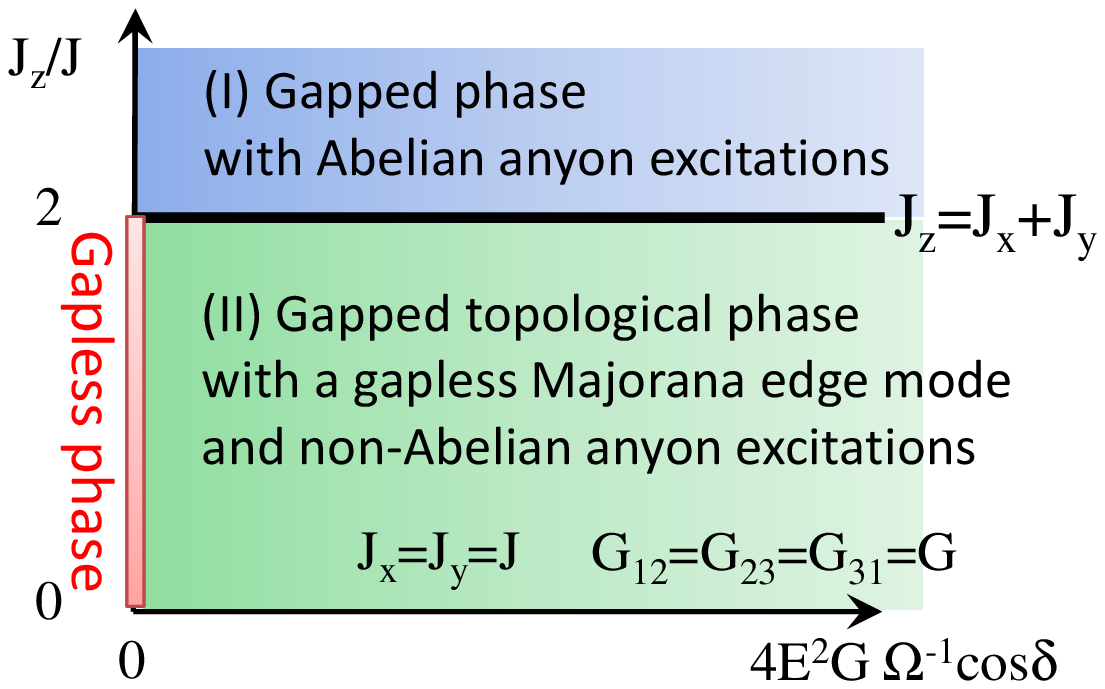}
\end{center}
\caption{(color online) Nonequilibrium phase diagram of the 
effective Hamiltonian~(\ref{eq:Effective}) in the space of 
$J_{x,y}=J$ and $G_{12,23,31}=G$. This diagram becomes accurate 
at $\Omega\gg |J_{x,y,z}|$.}
\label{fig:Phase}
\end{figure}

The phase diagram of the effective 
Hamiltonian~(\ref{eq:Effective}) is summarized in Fig.~\ref{fig:Phase}. 
%where we focus on the parameter space of $J_x=J_y=J$ and 
%$G_{12}=G_{23}=G_{31}$. 
Phase II corresponds to the laser-driven gapped phase with a gapless 
Majorana edge mode. It is possible to generate islands 
of the topological spin liquid state by applying spot laser 
as shown in Fig.~\ref{fig:Kitaev_Laser}(b). Their position can 
be changed by slowly moving the spot positions. 
On the line $J_z=2J$, a nonequilibrium phase transition occurs 
in which two Dirac points $K_{\rm a,f}$ merge at 
${\bol k}'=(0,2\pi/\sqrt{3})$.
% and the perturbation term 
%$h_{\bol k}^z$ with ${\bol k}={\bol k}'$ becomes zero. 

%%%%%%%%%%%%%%%%%%%%%%%%%%%%%%%%%%%%
%%%%%%%%%%%%%%%%%%%%%%%%%%%%%%%%%%%%

%%%%%%%%%%%%%%%%%%%%%%%%%%%%%%%%%%%%%%%%%%%%%%%%%%%%%%%%%%%%
%%%%%%%%%%%%%%%%%%%%%%%%%%%%%%%%%%%%%%%%%%%%%%%%%%%%%%%%%%%%
%%%%%%%%%%%%%%%%%%%%%%%%%%%%%%%%%%%%%%%%%%%%%%%%%%%%%%%%%%%%
%{\it Detection of edge mode} $-$
Finally, we shortly discuss the detection scheme for 
the gapless chiral edge mode. 
As in other topological states, 
the edge state can be observed through transport measurements. 
Thermal transport is appropriate since the present edge mode is chargeless. 
At sufficiently low temperatures $T$, the thermal conductance 
along an edge is~\cite{Kane,Cappelli,Kitaev}
\begin{eqnarray}
G_{\rm th} &=& \frac{\pi k_B^2}{12} T.
\label{eq:ThermalConductance}
\end{eqnarray}
We emphasize that $G_{\rm th}$ is half compared to IQHE
because the edge mode is formed by Majorana fermions.

%%%%%%%%%%%%%%%%%%%%%%%%%%%%%%%%%%%%%%%%%%%%%%%%%%%%%%%%%%%%
%%%%%%%%%%%%%%%%%%%%%%%%%%%%%%%%%%%%%%%%%%%%%%%%%%%%%%%%%%%%
%%%%%%%%%%%%%%%%%%%%%%%%%%%%%%%%%%%%%%%%%%%%%%%%%%%%%%%%%%%%
{\it Conclusions} $-$
We have studied laser-induced non-equilibrium states of the Kitaev model
making use of the Floquet theory. 
When EPL is applied, we showed that a 
topological spin-liquid state with a gapless Majorana edge mode 
is generated if the effect of magneto-striction ME coupling is considered. 
The laser-induced gap and the direction of the edge current can be 
controlled by changing the strength and helicity of laser. 
%THz or mid-infrared lasers is suitable 
%for experiments since $J_{x,y,z}$ are usually 
%within the THz energy scale. 

%%%%%%%%%%%%%%%%%%%%%%%%%%%%%%%%%%%%%%%%%%%%%%%%%%%%%%%%%%%%
%%%%%%%%%%%%%%%%%%%%%%%%%%%%%%%%%%%%%%%%%%%%%%%%%%%%%%%%%%%%
%%%%%%%%%%%%%%%%%%%%%%%%%%%%%%%%%%%%%%%%%%%%%%%%%%%%%%%%%%%%
\begin{acknowledgements}
MS deeply thanks Ken Shiozaki for discussions about 
topological superconductors. 
MS is also thankful to Ken Funo, Nobuo Furukawa, Masahito Mochizuki, 
and Sei Suzuki for several discussions. 
%This work is supported by Grants-in-Aid from JSPS, Grants No.
MS is supported by KAKENHI (Grants No. 25287088, 26870559), 
and TO by KAKENHI (Grant No. 23740260, 24224009). 
\end{acknowledgements}


\begin{thebibliography}{}
\bibitem{NoteGanichev}
Although chiral edge states were observed, 
it is highly likely that experiments~\cite{Karch10,Karch11} 
are in the classical limit of Hall effect.
\bibitem{note2}
When a laser is applied to the Kitaev model, 
$-{\bol E}\cdot{\bol P}_{\rm tot}$ acts on its quantum states. 
If we initially set $\hat I_{\bol r}=1$, the above operator does not 
induce any sign flip of $\hat I_{\bol r}$ because exchange couplings 
$\sigma^x_{\bol r}\sigma^x_{\bol r'}$, $\sigma^y_{\bol r}\sigma^y_{\bol r'}$, 
and $\sigma^z_{\bol r}\sigma^z_{\bol r'}$ in $\bol P_{\rm tot}$ 
are respectively proportional to $i\xi^a_{\bol r}\xi^b_{\bol r+\bol u_1}$, 
$-i\xi^a_{\bol r}\xi^b_{\bol r+\bol u_2}$ and 
$-i\hat I_{\bol r}\xi^a_{\bol r}\xi^b_{\bol r}$ 
[see Eq.~(\ref{eq:Kitaev_Majorana})].
\end{thebibliography}
\end{document}